\documentclass[%
 reprint,
superscriptaddress,
nolongbibliography,
 amsmath,amssymb,
 aps,
]{revtex4-2}

\usepackage{color}
\usepackage{graphicx}
\usepackage{dcolumn}
\usepackage{bm}
\usepackage[colorlinks= true,urlcolor=blue,linkcolor= blue,citecolor=blue,bookmarks=false,pdfstartview=]{hyperref}

\usepackage{color}

\begin{document}

\preprint{APS/123-QED}

\title{Low-temperature spin dynamics and absence of magnetic order in layered $\alpha$-RuI$_3$}%
\author{Hank C. H. Wu}
 \email{hank.wu@physics.ox.ac.uk}
\author{Benjamin M. Huddart}
\affiliation{Clarendon Laboratory, Department of Physics, University of Oxford, Parks Road, OX1 3PU, United Kingdom}%
\author{Francis L. Pratt}
\affiliation{ISIS Facility, Rutherford Appleton Laboratory, Chilton, Oxfordshire OX11 0QX, United Kingdom}%
\author{Danrui Ni}
\author{Robert J. Cava}
\affiliation{Department of Chemistry, Princeton University, Princeton, New Jersey 08544, United States}%
\author{Stephen J. Blundell}
 \email{stephen.blundell@physics.ox.ac.uk}
\affiliation{Clarendon Laboratory, Department of Physics, University of Oxford, Parks Road, OX1 3PU, United Kingdom}%

\date{\today}%

\begin{abstract}
The triangular-lattice system $\alpha$-RuI$_3$ is isostructural to the widely-studied $\alpha$-RuCl$_3$ compound which was identified as a potential Kitaev system but exhibits, instead of spin liquid behaviour, a magnetically ordered zig-zag ground state which sets in below 14~K.  Here we show experimentally that, in contrast, the spins in $\alpha$-RuI$_3$ remain dynamic down to at least 50~mK.  We study the spin dynamics using muon-spin relaxation methods and determine the presence of low-frequency fluctuations which are characteristic of a two-dimensional system.
\end{abstract}

\maketitle

Materials with magnetic ions decorating a layered honeycomb lattice are of great current interest due to the geometric frustration that is present when the magnetic ions are antiferromagnetically coupled~\cite{Balents2010}. In this context, $\alpha\text{-RuCl}_3$ has been widely studied as a potential quantum spin liquid (QSL) material.   The Ru$^{3+}$ ions in this compound form a layered honeycomb structure~\cite{Plumb2014} and this highlighted $\alpha\text{-RuCl}_3$ as a potential realisation of the exactly solvable Kitaev model~\cite{Kitaev2006}. The octahedral crystal field and spin-orbit coupling acting on the 4d$^5$ electrons produce a local effective $J = \frac{1}{2}$ ground state and lead to anisotropic interactions that are potentially described by the Kitaev model~\cite{Banerjee2016,Banerjee2017}.

However, $\alpha$-RuCl$_3$ orders magnetically at around 7~K to 14~K in zero-field, the temperature probably depending on stacking faults~\cite{Fletcher1967,Kobayashi1992,Johnson2015, Majumder2015,Sears2015,Lang2016}. It only enters a possible spin liquid state under an applied magnetic field at 10~T~\cite{Baek2017,Banerjee2018}. Meanwhile, $\alpha$-RuBr$_3$ was found also to possess a zigzag antiferromagnetic ground state with a higher transition temperature at 34~\text{K}~\cite{Imai2022,Choi2022}. The magnetic ordering in both compounds has been attributed to non-Kitaev interactions in the Hamiltonian, and both compounds are now thought to realise the more general $J$-$K$-$\Gamma$ model~\cite{Rau2014}, the Hamiltonian for which is given by
\begin{equation}
    \hat{H}=\sum_{\langle i,j\rangle}
    J {\bf S}_i\cdot{\bf S}_j
    + K S_i^\gamma S_j^\gamma
    + \Gamma \left(S_i^\alpha S_j^\beta
    + S_i^\beta S_j^\alpha \right),
\end{equation}
where $J$ is the Heisenberg exchange, $K$ is the Kitaev exchange, and $\Gamma$ the off-diagonal interaction.  On each bond, one spin direction is chosen as $\gamma$, with $\alpha$ and $\beta$ denoting the remaining directions. Although both the chloride and bromide salts exhibit magnetic order, the magnetic order in $\alpha$-RuBr$_3$ appears more robust.  The larger ionic radius of the Br$^-$ anion both expands the lattice and increases the covalency of the Ru--halide bond due to the up-floating of Br 4p bands compared with Cl 3p bands~\cite{Imai2022}, both of which effects tend to enhance the Kitaev interaction. 

The isostructural iodide $\alpha\text{-RuI}_3$ has only been recently synthesized and has shown promising potential to become another candidate QSL~\cite{Ni2022}. Magnetic susceptibility measurements of polycrystalline samples found $\alpha\text{-RuI}_3$ to be weakly paramagnetic and no magnetic ordering was observed down to $1.8$~K. The magnetisation data show a linear dependence on the external field at $200$~K, but at low temperature ($2$~K) there is an S-shaped feature which cannot be completely explained by a Brillouin function~\cite{Ni2022}. Heat capacity ($C_\text{p}$) measurements also found no indication of a phase transition between 0.35 and 10~K, but did identify a large $T$-linear term ($29.3 \text{ mJ} \text{ mol}^{-1} \text{ K}^{-2}$, larger than other layered insulating metal trihalides such as CrI$_3$, $1.17 \text{ mJ} \text{ mol}^{-1} \text{ K}^{-2}$, and in contrast with $\alpha\text{-RuCl}_3$ in which the $\gamma T$ term vanishes)~\cite{Ni2022}.

Resistivity measurements reveal metallic behaviour~\cite{Ni2022,Nawa2021}, consistent with density functional theory (DFT) calculations that identify metallic states near the Fermi surface~\cite{Ni2022, Nawa2021,Zhang2022}, and in contrast with the insulating nature of $\alpha$-RuCl$_3$.  However, it has been suggested that the metallic nature of the sample may be due to a grain boundary effect and might not be intrinsic~\cite{Kaib2022}, but recent resonant inelastic x-ray scattering data support the existence of a metallic state in $\alpha$-RuI$_3$ \cite{Gretarsson2024}. In this Letter, we report muon spin relaxation ($\mu$SR) data obtained on $\alpha$-RuI$_3$ and thereby demonstrate that the Ru$^{3+}$ ions in this compound do not form a magnetically ordered ground state down to 50~mK, demonstrating that this compound may realise a spin liquid ground state.
\begin{figure}
\hspace{-0.5cm}
\includegraphics[width=0.5\textwidth]{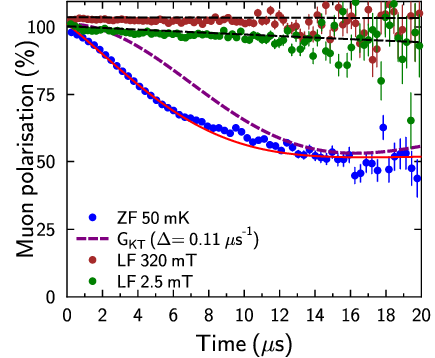}
\caption{Blue: ZF-$\mu$SR polarisation measured at 50~mK with error and a fitted line (red) by Eqn.~\ref{eqn:Asym_ZF}. Purple: A Kubo-Toyabe relaxation with $\Delta\approx 0.11~\mu s^{-1}$ as a comparison to a pure relaxation by nuclear moments. Green \& Brown: LF-$\mu$SR polarisation with 2.5~mT and 320~mT applied at 100~mK, with black dashed lines indicating the fit to an exponential decay function $e^{-\lambda t}$.}
\label{fig:RuI3_ZF} 
\end{figure}

The polycrystalline sample used in our experiment was prepared using a self-flux method (as described in~\cite{Ni2022}) and has been characterised using X-ray Diffraction. In a $\mu$SR experiment~\cite{Blundell2022} spin-polarized muons are implanted in a sample and interact with the local magnetic field. They subsequently decay, after an average time of 2.2~$\mu$s, and the angular distribution of the detected positrons, each of which is preferentially emitted in the direction of the muon spin at the time of decay, allows us to deduce the polarization of the muon-spin ensemble. We performed our zero-field (ZF) $\mu$SR spectroscopy using the HiFi instrument at the ISIS Pulsed Neutron and Muon Source with the sample loaded into a packet made of 25~$\mu$m silver foil, placed on a silver plate, and mounted in a dilution refrigerator~\cite{data}. The results, plotted in Fig.~\ref{fig:RuI3_ZF}, show a monotonic relaxation of the muon asymmetry at all temperatures, with no evidence of any precession signal that would indicate magnetic ordering. The asymmetry instead exhibits an initial Gaussian-like decay but becomes exponential at late times. This behavior can be crudely fitted by
\begin{equation}\label{eqn:Asym_ZF}
P(t)= P_\text{r}G_\text{KT}(t,\Delta)\, e^{-\lambda t} + P_\text{bg}
\end{equation}
where $P_\text{r}$ and $P_\text{bg}$ are the amplitudes relating to the relaxing component (due to muons implanted in the sample) and background component (due to muons implanted in the silver packet and plate). This can be interpreted by assigning the function $G_\text{KT}(t,\Delta)$, a Kubo-Toyabe relaxation function, to a distribution of static, but randomly-oriented, nuclear moments and assuming that the muon is also coupled to dynamical fluctuations in the field at the muon site which arise from fluctuations in the electronic Ru$^{3+}$ moments, giving rise to the weak exponential decay term $e^{-\lambda t}$. To illustrate the effect of this exponential, the pure Kubo-Toyabe function with field distribution width $\Delta = 0.11~\mu s^{-1}$ and no extra decay term is plotted as a dashed line in Fig.~\ref{fig:RuI3_ZF} (this value of $\Delta$ has been chosen to be consistent with the expected nuclear broadening at the calculated muon site, as will be discussed later). We found no possible fit to the data using just $G_\text{KT}(t,\Delta)$ which indicates that there is an additional relaxation process occurring in addition to the effect of static magnetic fields from nuclear moments. These ZF-$\mu$SR results therefore provide strong evidence that the sample does not undergo magnetic ordering down to 50~mK, in strong contrast to the spin precession signals that are obtained in ZF-$\mu$SR experiments on $\alpha$-RuCl$_3$~\cite{Lang2016} and $\alpha$-RuBr$_3$~\cite{Weinhold2023} below their antiferromagnetic ordering transitions.

\begin{figure}
\includegraphics[width=0.47\textwidth]{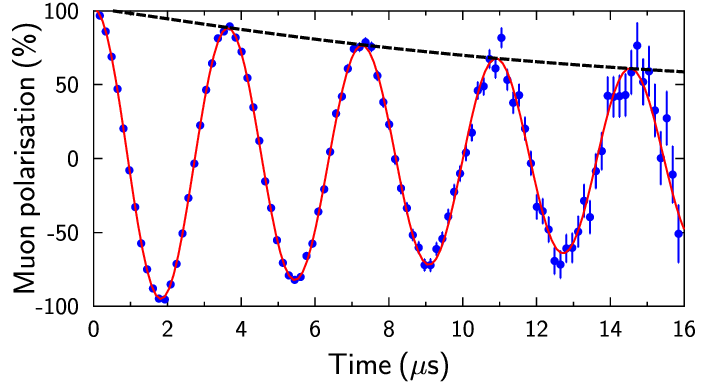}
\caption{wTF muon asymmetry (blue) at 61~mK with a 2.0~mT transverse field,
and the fitted line (red) using Eqn.~\ref{eqn:RuI3_wTF}. The dashed line (black) shows the relaxation envelope.}
\label{fig:RuI3_wTF}
\end{figure}

Further confirmation of this comes from weak transverse field (wTF) measurements.  Data taken with a B=2.0~mT transverse field at 61~mK exhibit a precession signal at the Larmor frequency with weak damping, as shown in Fig.~\ref{fig:RuI3_wTF}.By fitting the TF-$\mu$SR spectrum to a cosine function multiplied by a relaxation envelope, given by
\begin{equation}\label{eqn:RuI3_wTF}
    P(t) = (P_\text{r}e^{-\lambda t} + P_\text{bg})\cos(2\pi\nu t +\phi),
\end{equation}
and using the same relaxing asymmetry $P_r$ as in the ZF measurement (see Eqn.~\ref{eqn:Asym_ZF}), the precession frequency $\nu=\gamma_\mu B / (2\pi)$ is consistent with the applied field (where the muon gyromagnetic ratio $\gamma_\mu$ is given by
$\gamma_\mu=2\pi\times 135.5$~MHz\,T$^{-1}$, and 
$\lambda$ is found to be $0.074(2)~\mu$s$^{-1}$. 
We also fitted the spectrum with a stretched exponential decay $e^{(-\lambda t)^\beta}$ and obtained $\beta = 1.17(4)$ which is consistent with a Lorentzian field distribution  superimposed on top of the applied field, instead of a Gaussian one ($\beta = 2$). This reflects persistent spin dynamics and a fluctuating local field at the muon site.

\begin{figure}
\includegraphics[width=0.46\textwidth]{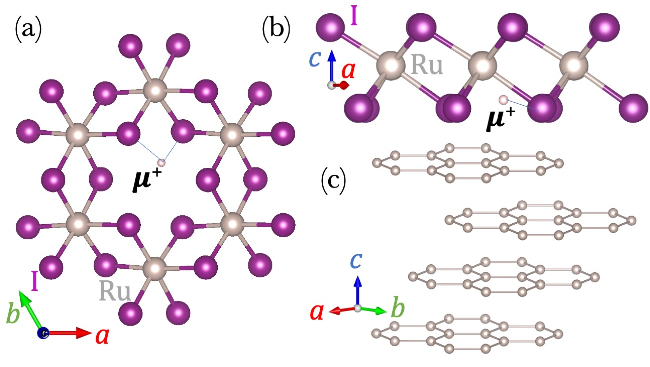}
\caption{Muon site for $\alpha$-RuI$_3$, viewed (a) along the $c$-axis, showing the structure of the layer, and (b) viewed along the $b$-axis, showing the layer edge-on. (c) The honeycomb Ru$^{3+}$ layers stacked along the $c$-axis.}
\label{fig:RuI3_muonsite}
\end{figure}

Using DFT (in particular, the DFT+$\mu$ method~\cite{Moller2013_FuF,Bernardini2013,Blundell2023}), it is possible to identify the muon site. Our calculations reveal a muon site in $\alpha$-RuI$_3$ at [0.544, 0.180, 0.286], within a RuI$_3$ layer, and close to the centre of a Ru-hexagon, as shown in 
Fig.~\ref{fig:RuI3_muonsite}(a,b). The site is in close proximity to two I$^{-}$ ions at distances of 1.91~\AA\, and 2.03~\AA\, from the muon, and with an I-$\mu$-I angle of 156.84\textdegree. The compound consists of stacks of RuI$_3$ hexagonal layers [see Fig~\ref{fig:RuI3_muonsite}(c)] and this site implies that the muon will be particularly sensitive to the moments within a single layer.  

To probe the nature of fluctuations in the local magnetic field at the muon site, we studied the effect of an applied longitudinal field (LF) at five temperatures from 0.1~K up to 40~K. The purpose of applying a longitudinal field is to quench the relaxation due to any static magnetic field from nearby nuclear moments and study the dynamical relaxation. In materials without electronic magnetism, a small applied field of 2.5~mT is sufficient to decouple the muon from the field produced by surrounding nuclear moments. Our results show that there is a significant recovery of muon polarization at 2.5~mT, which confirms that the muon experiences significant relaxation from nuclear moments. However, a small fraction of the relaxation in the form of $e^{-\lambda t}$ remains, as shown in Fig.~\ref{fig:RuI3_ZF}, which can be traced back to the effect of the fluctuating Ru$^{3+}$ electronic moments.

\begin{figure}
\hspace{-0.5cm}
\includegraphics[width=0.4\textwidth]{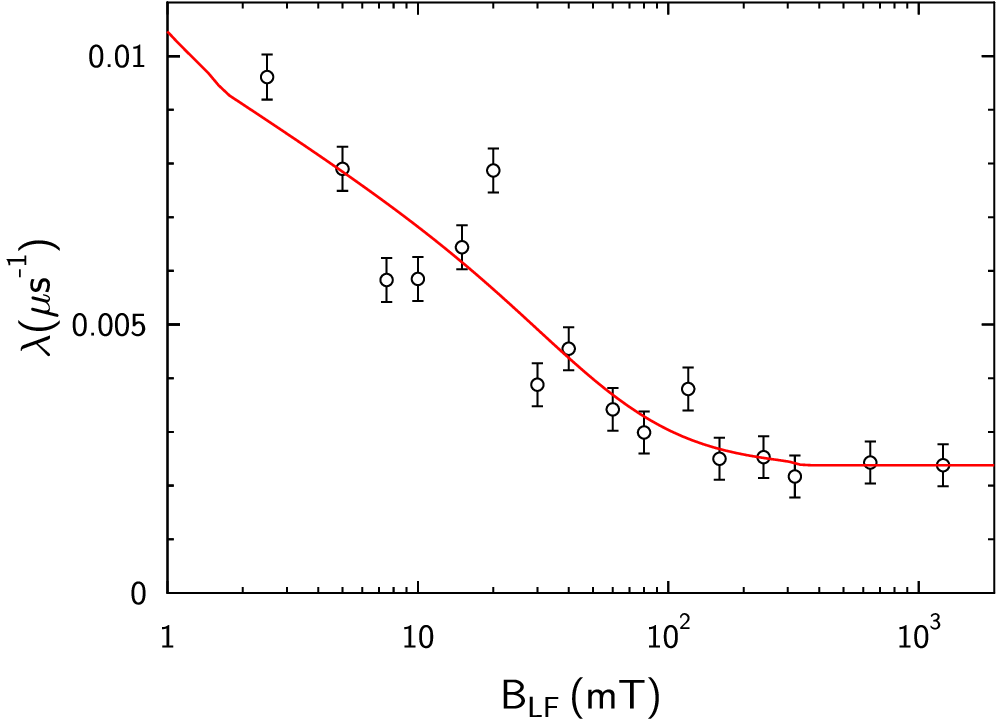}
\caption{Dynamical relaxation rate $\lambda$ as a function of applied longitudinal field $B_\text{LF}$ applied on $\alpha$-RuI$_3$ at 0.1~K.}
\label{fig:RuI3_LF_fit}
\end{figure}
The dynamical relaxation weakens with increasing longitudinal field. This is demonstrated by Fig.~\ref{fig:RuI3_LF_fit} which plots the relaxation rate $\lambda$ at 0.1~K as a function of applied longitudinal field for fields between 2.5~mT and 1.25~T. The plot could be interpreted as a combination of a field-dependent term and a field-independent contribution such that
\begin{equation}
    \lambda(B_\text{LF}) = \lambda_\text{2D}(B_\text{LF})+\lambda_0
\end{equation}
where $\lambda(B_\text{LF})$ refers to the diffusive spin excitation. Figure~\ref{fig:RuI3_LF_fit} also shows that a field of around 0.5~T is required to fully suppress the field-dependent spin diffusion in $\alpha$-RuI$_3$, in which case the remaining relaxation is then due to the effect of the localised spin excitation $\lambda_0$~\cite{Pratt2022}.

To study the spin correlations in $\alpha$-RuI$_3$, we fitted the relaxation rate $\lambda(B)$ at five temperatures against both the zero-dimensional Redfield model, where spins fluctuation is local and is not correlated to any of its neighbours, and a 2D spin diffusion model, as shown in Fig.~\ref{fig:RuI3_LF_fit}. The 2D model, which accounts for the field-dependent spin excitation term~\cite{Pratt2022, Pratt2023}, was consistently found to fit the LF dataset better than the zero-dimensional model. This is because the latter results in a relaxation rate plateau at low fields, but experimentally the relaxation rate is found to increase continuously even at very low fields such as 2.5~mT, as shown in Fig.~\ref{fig:RuI3_LF_fit}.

\begin{figure}
\hspace{-0.5cm}
\includegraphics[width=0.45\textwidth]{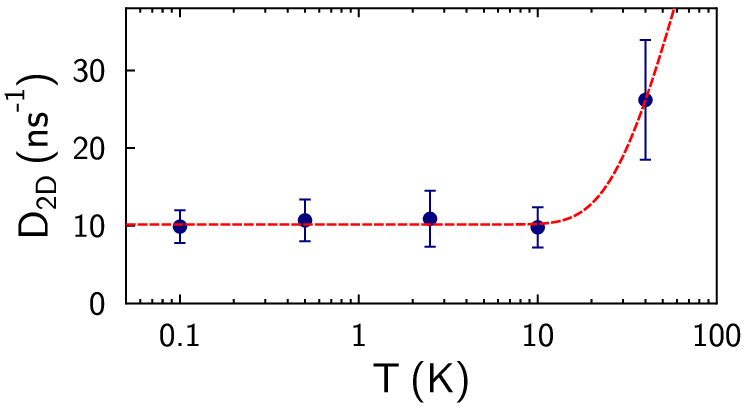}
\caption{The fast diffusion rate $D_\text{2D}$ of the 2D spin diffusion model extracted by fits in Fig.~\ref{fig:RuI3_LF_fit} at five temperatures from 0.1~K to 10~K.}
\label{fig:RuI3_D2d}
\end{figure}

The spin diffusion rate $D_\text{2D}$ is plotted as a function of temperature T in Fig.~\ref{fig:RuI3_D2d}, stays around 10~$\text{ns}^{-1}$ between 0.1~K and 10~K, but shoots up to 26~$\text{ns}^{-1}$ at 40~K (the line in Fig.~\ref{fig:RuI3_D2d} is a guide to the eye, but assumes that there is a temperature independent and an activated temperature dependent process contributing in parallel). The rising diffusion rate corresponds to a falling entanglement length and is consistent with the scenario where in the high-$T$ classical regime the Ru$^{3+}$ moments are correlated to only their nearest neighbours, whereas at low-$T$ (in the quantum regime) the QSL state has a longer entanglement range between spin pairs~\cite{Pratt2023}. We note that the reported Curie-Weiss temperature (15.4~K \cite{Kaib2022}) lies within the crossover regime
between 10~K and 40~K.

The LF-$\mu$SR analysis therefore demonstrates that $\alpha$-RuI$_3$ shows the characteristic field dependence of a 2D spin diffusion model, so that there is a fast diffusion within the $ab$ plane (deduced by the crystal symmetry) and slow diffusion along $c$. This interpretation supports the idea that there are antiferromagnetic spin correlations among the weak Ru$^{3+}$ moments beyond the nearest neighbours, and also that there is a crossover from the quantum to classical regime between 10~K and 40~K, based on the fitted spin diffusion rate $D_\text{2D}$.

We have estimated the relaxation due to the magnetic field resulting from 
randomly-oriented nuclear moments at the muon site identified by the DFT$+\mu$ method, as well as a couple of other less likely, but possible, muon sites (which have higher energy).  These estimates imply a Kubo-Toyabe relaxation with a value of  $\Delta$ (see Eq.~\ref{eqn:Asym_ZF}) between 0.03~$\mu$s$^{-1}$ to 0.08~$\mu$s$^{-1}$, since the shortest $\mu$-I bond ranges from 1.67~\AA\, to 1.93~\AA\ and the effect is mostly dominated by the distance to the nearest I$^{-}$ ion(s) as the Van Vleck sum goes as $\sigma^2 \propto \sum_i \frac{\mu_{i,I}}{r_i^6}$ (and is dominated by the spin $I=\frac{5}{2}$ iodine nuclei, which each have a moment 7.5 times larger than that of ruthenium nuclei on average, considering all relevant isotopes with their natural abundances). Fitting the 50~mK ZF-$\mu$SR data to a Kubo-Toyabe function together with an exponential relaxation to account for fluctuating moments gives a fitted value of $\Delta= 0.11~\mu$s$^{-1}$, confirming the muon's proximity to the nearest I$^{-}$ ion. We have also estimated the magnitude of the static field that would result from antiferromagnetic order, as is observed in the case of $\alpha$-RuCl$_3$ and $\alpha$-RuBr$_3$.  We find that all conceivable muon sites should have a dipolar field of at least 1.5~mT (or muon precession frequency of 0.2~MHz) if Ru$^{3+}$ moments are ordered antiferromagnetically with a moment size of 0.1~$\mu_\text{B}$ each. This effect would correspond to an oscillation period of less than 5~$\mu$s$^{-1}$, and this effect is certainly not observed in our ZF-$\mu$SR data (see Fig.~\ref{fig:RuI3_ZF}). In addition, should there be a magnetic order the local static field would have damped the 2~mT wTF-$\mu$SR measurements very effectively (see Fig.~\ref{fig:RuI3_wTF}), which they do not. Therefore we can conclusively rule out static order for $\alpha$-RuI$_3$ down to 50~mK, in stark contrast to what has been observed with $\mu$SR for
both $\alpha$-RuCl$_3$~\cite{Lang2016} and $\alpha$-RuBr$_3$~\cite{Weinhold2023}.

Due to the more delocalised 4d orbitals in RuI$_3$ than in RuCl$_3$~\cite{Kaib2022}, d-p hybridisation is understood to be stronger in RuI$_3$~\cite{Kim2021}. It was reported that while the zigzag ordered state was calculated to be the energetically favourable state in RuCl$_3$ and in RuBr$_3$, all magnetic states considered for RuI$_3$ were very similar in energy, which has motivated the search for a QSL state in this compound~\cite{Kaib2022}. DFT calculations performed on RuI$_3$ have predicted the magnitude of the Ru$^{3+}$ moments to be very small, of the order of 0.1~$\mu_B$~\cite{Liu2023, Kaib2022}, and RIXS spectra imply that the low-temperature magnetism can be described using $J=1/2$ pseudospins, even though the compound is metallic~\cite{Gretarsson2024}. This is all consistent with susceptibility measurements~\cite{Ni2022} and our $\mu$SR results.

{\it Conclusion:} Therefore, through zero-field and weak transverse-field $\mu$SR measurements, we are able to demonstrate the absence of magnetic order in $\alpha-$RuI$_3$ down to 50~mK and that its local magnetism can be explained in terms of the effect of both the static nuclear moments and the fluctuating electronic moments. Because these fluctuating moments are small in magnitude, their presence is only directly observable at late times, from $t = 6~\mu s^{-1}$ onwards, when the nuclear moments-led relaxation fades out. Despite the small size of the Ru$^{3+}$ moments, their fluctuating dynamics are revealed by using an applied longitudinal field that suppresses the relaxation from the nuclear moments. Our analysis shows that the nature of the spin correlations is consistent with a 2D spin diffusion model which has a transition between a quantum and a classical regime between 10~K and 40~K. The results therefore demonstrate the strong likelihood that $\alpha-$RuI$_3$ hosts a quantum spin liquid state.

{\it Acknowledgments:}
The $\mu$SR experiments were carried out at the ISIS neutron and muon source, UK, and were supported by beam-time allocation RB2310371. We acknowledge the support by the ISIS Neutron and Muon Source, the scholarship funding by the Croucher Foundation, and the UK Research and Innovation (UKRI) under the UK government’s Horizon Europe funding guarantee [Grant No. EP/X025861/1].

\bibliography{library.bib}

\end{document}